\documentclass[preprint,journal]{vgtc}       

\ifpdf
  \pdfoutput=1\relax                   
  \usepackage{graphicx}                
  \DeclareGraphicsExtensions{.pdf,.png,.jpg,.jpeg} 
\else
  \usepackage{graphicx}                
  \DeclareGraphicsExtensions{.eps}     
\fi%

\graphicspath{{figures/}{pictures/}{images/}{./}} 

\usepackage{microtype}                 
\PassOptionsToPackage{warn}{textcomp}  
\usepackage{textcomp}                  
\usepackage{mathptmx}                  
\usepackage{times}                     
\usepackage{cite}                      
\usepackage{tabu}                      
\usepackage{booktabs}                  
\usepackage{soul}

\title{ConceptExplainer: Interactive Explanation for Deep Neural Networks from a Concept Perspective}

\author{Jinbin Huang, Aditi Mishra, Bum-Chul Kwon, Chris Bryan}
\authorfooter{
\item
 {Jinbin Huang, Aditi Mishra and Chris Bryan} is with Arizona State University. E-mail: \{jhuan196, amishr45, cbryan16\}@asu.edu
\item
 Bum-Chul Kwon is with IBM research. E-mail: bumchul.kwon@us.ibm.com.
}

\newcommand{\name}{\textsc{ConceptExplainer}}

\abstract{Traditional deep learning interpretability methods which are suitable for non-expert users cannot explain network behaviors at the global level and are inflexible at providing fine-grained explanations. As a solution, concept-based explanations are gaining attention due to their human intuitiveness and their flexibility to describe both global and local model behaviors. Concepts are groups of similarly meaningful pixels that express a notion, embedded within the network's latent space and have primarily been hand-generated, but have recently been discovered by automated approaches. Unfortunately, the magnitude and diversity of discovered concepts makes it difficult for non-experts to navigate and make sense of the concept space, and lack of easy-to-use software also makes concept explanations inaccessible to many non-expert users. Visual analytics can serve a valuable role in bridging these gaps by enabling structured navigation and exploration of the concept space to provide concept-based insights of model behavior to users. To this end, we design, develop, and validate \name{}, a visual analytics system that enables non-expert users to interactively probe and explore the concept space to explain model behavior at the instance/class/global level. The system was developed via iterative prototyping to address a number of design challenges that non-experts face in interpreting the behavior of deep learning models. Via a rigorous user study, we validate how \name{} supports these challenges. Likewise, we conduct a series of usage scenarios to demonstrate how the system supports the interactive analysis of model behavior across a variety of tasks and explanation granularities, such as identifying concepts that are important to classification, identifying bias in training data, and understanding how concepts can be shared across diverse and seemingly dissimilar classes.

} 


\CCScatlist{ 
 \CCScat{K.6.1}{Management of Computing and Information Systems}%
{Project and People Management}{Life Cycle};
 \CCScat{K.7.m}{The Computing Profession}{Miscellaneous}{Ethics}
}

\teaser{
  \centering
  \includegraphics[width=\linewidth]{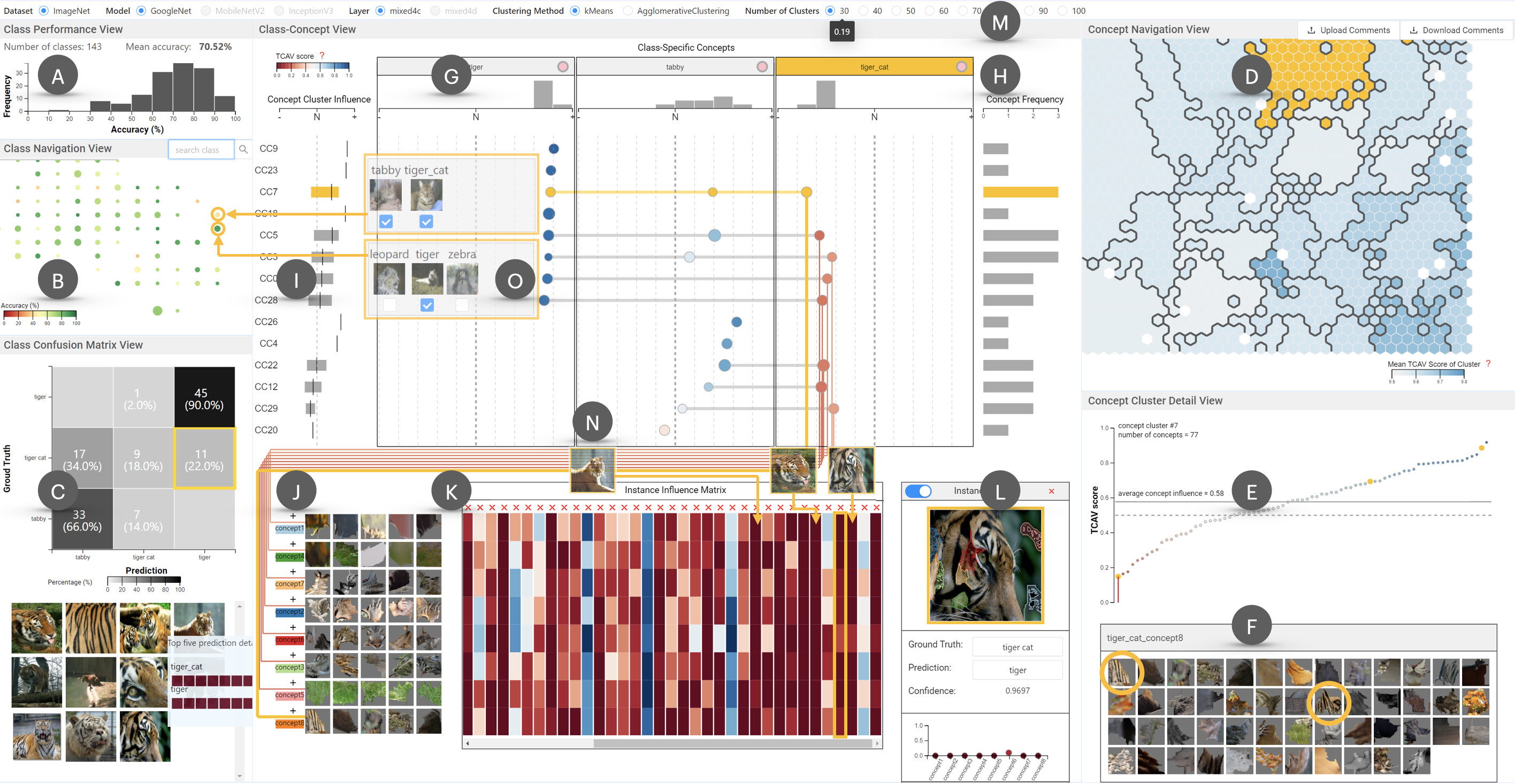}
  \caption{\name{} provides interactive concept-based explanations on deep neural networks (DNNs) at global/class/instance levels via a coordinated visual analytics interface connected to a backend processing pipeline. A full description of the interface design is given in Sect.~\ref{sec:frontend}; this image also shows actions taken during the usage scenario described in Sect.~\ref{sec:usage_scenario_2}. i) Class Navigation Panel: A, B, C; ii) Concept Navigation Panel: D, E, F; iii) Class-Concept Panel: G, H, I, J, K, L; iv) Header Bar Panel: M.  
  }
  \label{fig:teaser}
}

\vgtcinsertpkg

\begin{document}

\firstsection{Introduction}

\maketitle

Deep learning has led to tremendous advances across a number of fields, including natural language processing, speech recognition, medical applications, computer vision, and autonomous vehicles~\cite{dong2021survey}. Unfortunately, the complexity and inherent opaqueness of neural networks imposes significant difficulty in understanding the behavior and inner workings of models~\cite{dargan2020survey}. As deep learning is increasingly employed in today's society, it is important that people (including those who are non-experts in AI) are able to understand and interpret the predictions made by AI models~\cite{xie2020explainable} --- the explainable AI (XAI) subfield supports the development of tools and techniques to support this process~\cite{xu2019explainable}.

Generally, interpretability of trained neural networks is achieved either by (1) revealing the neural architecture, and how signals flow through the pipeline, or (2) using explainable substitutions to approximate the neural network's mental model on a task~\cite{gilpin2018explaining}. While the first option is straightforward (i.e., making transparent the inner logic and algorithmic functions of the model) it requires users that are equipped with deep learning expertise, and is thus inappropriate as a paradigm for non-experts~\cite{liao2020questioning}. In contrast, the latter option fits users with little deep learning expertise, but it must be carefully employed as it comes with inherent information loss and potentially false reflection~\cite{dovsilovic2018explainable}.
One interpretability approach that is suitable for non-experts is to demonstrate how input perturbation affects model outputs. In image classification, saliency and activation maps~\cite{selvaraju2017grad, zhou2016learning, saha2020role, simonyan2013deep, fong2017interpretable} measure the importance of the input features (i.e., image pixels) when classifying an image. The drawback is that these approaches only provide \textit{local} or sample-level explanations for individual images. Cross-grained insights which can help users understand \textit{class} or \textit{global}-level model behaviors --- e.g., measuring the influence that ``ear'' and ``fur'' have on predicting a classification of a ``cat'' --- is not possible~\cite{tcav}. Such understanding can be important, however, for non-expert users in the interpretability process, as it allows them to compare their own mental models of how classification should work to the neural network's mental model~\cite{kambhampati2021symbols}.
To facilitate such explanations, concept-based methods have recently been introduced by the AI community ~\cite{tcav, ace, szegedy2016rethinking, yeh2020completeness}.
For image classification, a concept refers to a group of pixels in a sample that represents an important part of the class object within the image~\cite{ace}. To be considered as a concept for a class, the concept needs to be easily understandable to human users, consistent within each concept but separable from other concepts, and necessary for the prediction of the class. Importantly, concepts can be used to generate explanations at the different levels of granularity.
At a high level, we can compute the concepts that are influential to classes (e.g., ``ear'', ``fur'', ``tail'' are likely important to the ``cat'' class). Locally (for an instance in a class), we can measure how much each concept influences the correct or incorrect classification of the class.
When concepts and classes are aggregated together, there is the potential to form a global perspective of the neural network's mental model, to understand the large-scale behavior based on what the network has learned.
Although concepts are appropriate for use as an interpretability technique for non-experts, there has been little broad application of this technique. This is likely because of two shared complexities that make concept generation and presentation a difficult exercise: (1)~The ``normal'' procedure to create concepts is by hand (i.e., manually). To automatically generate the massive concept space that covers an entire neural network is a non-trivial exercise. (2)~Structuring, navigating, and probing such a concept space in a way that supports different (global and local) explanation granularities is likewise non-trivial. 
To address these two challenges, we propose \name{}. \name{} consists of two primary software components: (1) a backend pipeline interactively (and automatically) generates and structures a concept space for a neural network, and (2) a coordinated, visualization-driven frontend lets users interactively explore and probe the concept space. \textbf{\name{} represents the first interactive visualization system designed for non-expert users that supports concept-based explanations of deep learning models at different levels of granularity.} At the global level,  \textsc{ConceptExplainer} can reveal which concepts are broadly influencing the neural network's decision making. At the class level, users can visualize which concepts are influential for that class, and can explore the overlaps of influential concepts which are shared between classes. At the instance level, individual images can be reviewed to see how present/non-present concepts affect the model's correct/incorrect predictions.
In developing and evaluating \name{}, we make the following research contributions: 
\textbf{(1) Identifying challenges and goals.} Based on an analysis of challenges in XAI for non-experts, we identify salient \textbf{design challenges} and \textbf{design goals} which are important for concept-based explanation for non-expert users, particularly in an interactive visualization context.
\textbf{(2) A novel visual analytics tool for concept-based explanation.} To support the ideated challenges and goals, we design and implement \name{}, an interactive system that automatically generates a concept space for an image classification neural network, and supports non-expert users to explore and probe concept-based explanations at multiple levels of granularity. \name{} is designed based on an iterative prototyping process, formatively validated through multiple evaluations (usage scenarios and a user study), and the codebase is open-sourced to promote adoption and replication.
\textbf{(3) Empirical findings and generalizable takeaways.} Based on the process of designing, implementing, and validating \name{}, we discuss several lessons and takeaways on the topic of how visualization-driven concept-based explanation can support XAI tasks, such as identifying issues (or biases) in data samples, and how tools like \name{} can be extended in the future (such as supporting expert users).

\section{Related Work}
\label{sec:related_work}

\subsection{Deep Learning Interpretability using Concepts}
There are two primary approaches for using concepts to improve deep learning interpretability: (1) training inherently interpretable models with concept-based constraints. (2) constructing post-hoc explanations based on concepts. 

Regarding the first approach, Koh et al.~\cite{cbm} proposed Concept Bottleneck Model (CBM), which restricts neural networks to behave in an auto-encoder manner where they first map inputs to human-interpretable concepts and then predict based on those concepts. Chen et al. ~\cite{cw} proposed Concept Whitening (CW) layer as a substitution for the normalization layer (i.e., batch normalization) to achieve higher interpretablity. While these methods employ concepts to offer embedded interpretablity with no extra dependency, they do not provide ways to gain insights into a trained model. Our work differs from this line of research as we seek to interpret fixed models with no alteration.

For the second approach,
TCAV~\cite{tcav} provides a means to computationally define a concept and calculate its class-specific influence on model predictions (called the TCAV score). Building upon TCAV, ACE~\cite{ace} is a technique to automatically extract influential class-specific concepts from a dataset. For a formal description of ACE and TCAV, see Sect.~\ref{sec:backend}, where we describe how both techniques are incorporated into \textsc{ConceptExplainer}'s backend to automatically define and extract concepts.

Similarly, Ge at al.~\cite{vrx} proposed the Visual Reasoning Explanation (VRX) framework that answers interpretability questions such as \textit{Why?} and \textit{Why not?} ~\cite{liao2020questioning, mishra2021policy} from a concept perspective by extracting and organizing class-specific concepts using trained structural concept graphs (SCGs) based on pairwise concept relationships. Though VRX provides insightful instance analysis, it requires training an SCG for each class. This makes VRX difficult to implement within interactive scenarios (i.e., as a part of a user interface) due to computational cost.

\subsection{Visual Analytics in Concept-based Interpretability}
A large amount of neural network interpretability research in the visualization community has utilized (non-concept-based) conventional methods, including visualizing features, saliency and activation maps, and visualizing the model neurons or architectures (e.g.,~\cite{boggust2022embedding, sivaraman2022emblaze, park_vatun_2021}); for more information, there are several recent surveys that discuss XAI from a visualization perspective~\cite{hohman2018visual, zhang2018visual, yuan2021survey}. For research that employs visualization to support concept-based tasks, there are two recent systems that are highly relevant to \name{}:

\textsc{ConceptExtract}~\cite{zhao2021human} utilizes visualization within a human-in-the-loop pipeline for concept extraction and fine-tuning. The system generates initial concept segments using ACE; the user then interactively refines the set of concept images. The user can also participate in refining neural network-based binary concept classifiers by interactively supplying labels. The intent of the system is to counter the potential drawbacks of automatic concept extraction, such as human incomprehensible patches, by incorporating human oversight and manual labeling, with the ultimate goal of enabling efficient handcrafting of high-quality concepts.

\textsc{NeuroCartography}~\cite{nc} aims to interactively reveal neuron-concept relations by grouping a model's similarly-activated neurons in the same latent layer by the same set of data instances. The corresponding set of data instances is viewed as a set of concepts. Concepts discovered in this way are layer-dependent. They evolve as layers go deeper. The system is thus capable of letting users analyze concept evolution in the context of the neural architecture, which is potentially beneficial to expert users (e.g., model developers) in debugging.

While these two systems combine concept-based explanations and visual analytics, both differ from \name{} in two important ways: (1) They are designed for expert users; in contrast, we focus on AI non-experts. (2) They support different tasks (creating/refining concepts, and revealing concept/neuron relations); our system is designed to probe and explore concepts as a way to understand model behavior. To the best of our knowledge, \name{} is the first interactive visual analytics system designed for non-experts that employs concept as a means of explanation.

\subsection{Deep Learning Interpretability for Non-Expert Users}

The popularity of deep learning in today's society has led to increased demands for XAI techniques that are friendly to non-expert users ~\cite{mohseni2021multidisciplinary}. Desiderata for this type of explanation includes faithfulness, consistency with prior beliefs and generality~\cite{carvalho2019machine}. Liao et al. ~\cite{liao2020questioning} proposed a set of design principles for end users, targeting explanation system in form of a question bank, where global explanations for a model's decision making and local explanations for a particular decision on an instance were ranked as top interests. Similarly, Hohman et al. ~\cite{hohman2018visual}, in their categorization of end users (which include model developers/builders, model users, and non-experts) suggest that model users who apply deep learning models to their domain tasks mostly need tools that support exploration of model behaviors on local neighborhoods and at global scale. Concept-based explanations are a good fit for these needs, as they require minimal prior knowledge about neural networks (e.g., it is not necessary to know technical concepts like backpropagation) while remaining intuitive and accessible. \name{} supports probing and exploring of concepts at multiple levels to support interpretability tasks for non-experts.

\section{Design Challenges and Goals}

Similar to prior visualization design studies in XAI ~\cite{hohman2018visual, hohman2019s, samek2021explaining}, we identify a salient set of design challenges (C1--C4) which are important for concept-based explanation for non-expert users, based on reviewing of state-of-the-art publications that discuss challenges in XAI, concept-based explanations, and AI explanations for non-experts (see Section~\ref{sec:related_work}). We distill these into a set of four design goals (G1--G4), which we use to guide \name{}'s development. 

\begin{figure}[h]
    \centering
    \includegraphics[width=\linewidth]{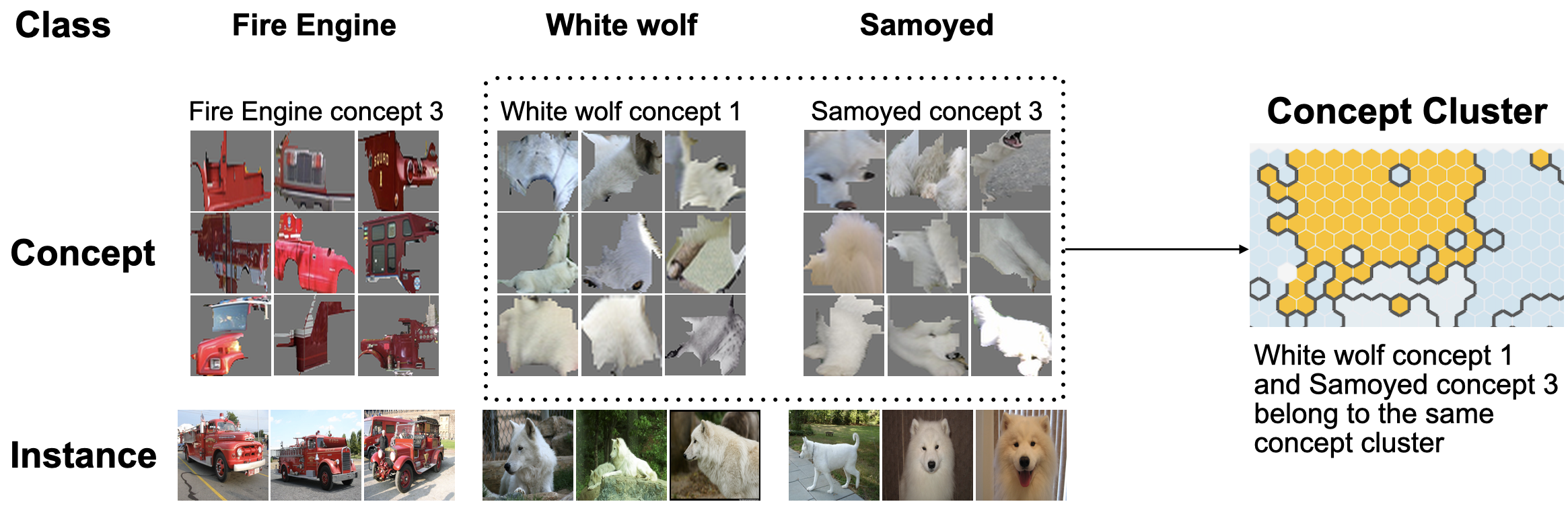}
    \caption{An example of how classes, concepts, and instances are related. Using our concept extraction and clustering pipeline (described in Sect.~\ref{sec:backend}, we show an example concept for each of the fire engine, white wolf, and Samoyed classes. }
    \label{fig:conceptDiagram}
\end{figure}

\begin{figure*}[h]
    \centering
    \includegraphics[width=\textwidth]{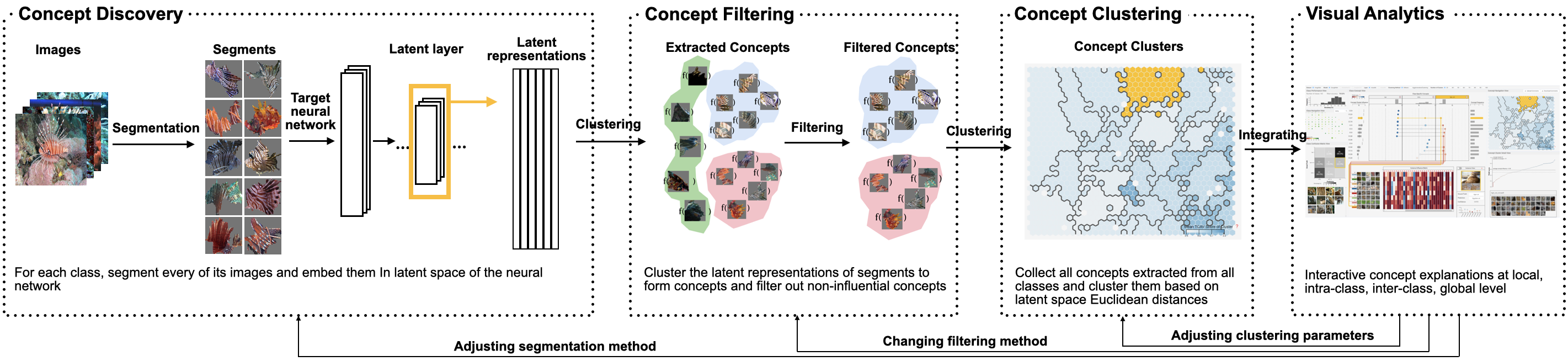}
    \caption{The workflow for \name{}. In the \textit{concept discovery} stage, class-specific images are segmented, embedded in latent space and clustered to form class-specific concepts. In the \textit{concept filtering} stage, non-influential concepts are filtered based on t-test on their TCAV scores. Concepts are then clustered in the \textit{concept clustering} stage. We employ \textit{visual analytics} in the frontend interface to let users explore and probe concepts at the local, class, and global levels; users can also interactively update parameters on the backend stages.}
    \label{fig:workflow}
    \vspace{-.5cm}
\end{figure*}

\subsection{Design Challenges}
\label{sec:design_challenges}

\textbf{(C1) Extracting human-understandable concepts for classes.}
There is increasing evidence that neural networks predict based on a combination of the concepts present in instances. ~\cite{olah2020overview, duggal2020rest, he2017channel, jaderberg2014speeding, wen2016learning} It is helpful for the user to understand neural network behavior by demonstrating what concepts a neural network relies on for predictions and the extent of concept influence on predictions. For example, when the neural network classifies a \textit{tiger}, is the ``bushy background'' concept more influential or the ``tiger stripe''? By extracting influential concepts for each class, the user can intuitively understand how the network understands each class and verify if it works in a sensible way. However, discovering concepts and measuring concept influences can be difficult as noise is inevitable and there are different ways to measure influence~\cite{zhao2021human}. In our work, we adopt the ACE method for concept extraction and TCAV method for measuring concept influence, as they are the most widespread approaches for concept-based explanations. We also propose a concept clustering process after concept discovery to structure the concept space and facilitate concept navigation.

\textbf{(C2) Multi-scale concept visualization for large datasets.} System scalability has gained increasing attention in the visualization community. The issue becomes salient when visual interfaces for XAI are concerned because models to be explained are trained on large datasets. Concept explanations also face this issue because the number of concepts discovered grows linearly with the number of classes in the dataset. To demonstrate the concept space and enable users to gain insights from navigating the space, we need more than a naive navigation mechanism which can cause the user to be lost in the deluge of concept information without a clear navigational goal. Apart from that, the data structure of concepts -- quantitative influences and images -- requires specifically designed visualizations to present.

\textbf{(C3) Revealing conceptual overlap between classes.}
The conceptual overlap (i.e., common influential features) between classes can reveal to the user why one class might be easily misclassified as another (e.g., Samoyeds classified as wolfs due to the influence of ``snowy background'' concept). Among a group of classes, understanding what concepts are shared (or not shared) can help discover features that cause confusion~\cite{goyal2019explaining} and/or serve as the unique ``signature'' of a class in the neural network's mental model.
For example, Fig.~\ref{fig:conceptDiagram}, shows a network has learned a ``white fur'' concept from both the \textit{white wolf} and \textit{Samoyed} classes, which provides an idea about the commonalities the network sees between these classes. In contrast, the lack of overlap between \textit{fire engine} and \textit{white wolf} also suggests that the network discriminates animals from vehicles.

\textbf{(C4) Balancing global explanations and local explanations.}
Global explanations without much detail are generally easier to understand. However, since global explanations are a summary of local behavior solely revealing global explanations might lead to unfaithful interpretations. To ensure faithfulness of the explanations we need to show information at both global and local levels and integrate the analysis process at both levels in an organic workflow \cite{ribera2019can}. The user benefits from having a holistic view of neural network behavior at various levels (see Fig. ~\ref{fig:conceptDiagram}) in a consistent concept-based framework so they can (1) recognize contradictory explanations (2) verify their hypothesis at different levels.

\subsection{Design Goals}
\label{sec:design_goals}

Based on the design challenges C1--C4, we identify four design goals for the \name{} system to support interactive concept-based explanations of neural networks' behavior for non-experts. Roughly, these goals can be ordered in terms of their granularity: at the global, class, and instance levels of analysis and explanation.

\textbf{(G1) Navigating the global concept space.} We aim to facilitate navigation in the concept space of a large dataset (in our case, we use ImageNet, which contains 1.2 M images for 1000 classes, see Sect.~\ref{sec:dataset_and_model}) (C2). The methodology we use is transferable to other image datasets as long as a trained deep learning model is available. To discover the concept space while preventing noise (C1), we first use ACE algorithm to extract influential concepts from the dataset. To facilitate structured concept navigation we cluster the extracted concepts into multiple concept clusters. The clustering of concepts makes the navigation easier -- the user can start their navigation by peeking at various concept clusters to pick up the one they want to explore and get into the cluster to see concepts inside. By gradually understanding the concept clusters the user can understand the concept space. The transition from concept cluster to individual concept reinforces an overview-first-detail-on-demand workflow. To help the user keep track of their investigation process in the concept space and understand the concept clusters by labeling them (taking notes of them) in a user-driven way, annotation functionality should also be available in the system.

\textbf{(G2) Supporting intra-class concept analysis} For a class, the discovered, influential concepts can be represented as a collection of image patches that have similar high dimensional activations. These influential concepts intuitively reveal how the model understands a class. The system needs to present these concepts such that the user can understand what they are and the extent of their influence (C2, C3). Users can even inspect the most influential concepts to see if the network has learned sensible relations (e.g., noisy concepts should be less influential than relevant concepts).

\textbf{(G3) Supporting inter-class concept analysis.}
Understanding conceptual commonalities between classes can help users to understand how the network perceives similar/different classes and pinpoint concept-based root causes for misclassifications. To this end, we seek to demonstrate the links between similar concepts of different classes (C3). It is necessary that inter-class concept analysis go beyond ``comparing two classes,'' as several classes may share commonalities of interest and the user may want to understand how the neural network considers them comprehensively.

\textbf{(G4) Supporting instance analysis.} Only explaining neural network behavior from a global or class level might lead the user to overlook important details, resulting in misinterpretations of how the network should work. To enable detailed inspection, while not overwhelming by revealing too many details, concept influences can be measured for individual instances. Such options keep the analytical framework unified: the user does not need other tools for instance explanations/analysis. In addition, instance-level analysis also helps users investigate edge cases (``\textit{Why did the model misclassify this image?}'') by listing influential concepts of the image.

\section{System Design}

Based on the design goals G1--G4, we develop \name{}, an interactive system that provides non-expert users with the ability to probe and analyze concept-based explanations at both the global and local levels. As shown in Fig.~\ref{fig:workflow}, the system consists of an integrated backend and frontend. In the \textbf{backend}, we design a processing pipeline that leverages concept-extraction methods (specifically, TCAV and ACE) to automatically extract and organize concepts for a given image dataset. Extracted concepts are organized into meaningful clusters to facilitate interactive concept navigation and concept overlap inspection. The \textbf{frontend} interface (Fig.~\ref{fig:teaser}) consists of four coordinated panels: (1) the header (M) contains controls widgets, including for manipulating the backend processing (setting parameters for the TCAV, ACE, and clustering methods), (2) the left panel  (A -- C) provides overview, navigation and analysis for classes, (3) the right panel (D -- F) provides a clutter-free locality-retaining overview for the concept space and navigation mechanisms reinforcing the overview-first-detail-on-demand mantra. (4) the central panel (G -- L) provides support for class/instance level analysis -- an overview of class-specific concepts and inter-class concept links combined with a detailed concept inspection view account for both class and instance level model behaviors.

\subsection{Iterative Prototyping Process}
\label{sec:iterative_prototyping}

To facilitate our design process, we employed an iterative prototyping methodology~\cite{nielsen1993iterative}. Over an approximately six months period, we sketched and prototyped a number of user interface and interaction designs. Designs were holistically reviewed and discussed by the project team, with additional feedback solicited from XAI researchers who work on interpretability techniques for non-expert users. Designs deemed suboptimal were either discarded or refactored (Fig.~\ref{fig:prototypes} in the Appendix shows two examples of ``old'' interfaces), while well-received designs were iteratively refined to develop the current system version. As an example of this process' impact, we adopted a three-panel layout in the interface (as shown in Fig.~\ref{fig:teaser} and described in Sect.~\ref{sec:frontend}). The left \textbf{class navigation panel} supports class-based analysis and navigation; in parallel, the right \textbf{concept-navigation panel} supports concept-based analysis and navigation. Within each of these panels, users can transit fluently from global perspectives of the model (i.e., visualizing all classes/concepts together) down to analyzing individual instances. Classes, concepts, and instances are synthesized together in the center \textbf{class-concept panel}.

\subsection{Dataset and Model}
\label{sec:dataset_and_model}

To demonstrate a real-world application throughout the rest of this paper, we use the ImageNet dataset~\cite{ImageNet}, which is a well-known image dataset consisting of 1.2M training instances. As a model, we employ GoogleNet~\cite{GoogleNet}, a convolutional neural network consisting of 22 layers. GoogleNet provides a pre-trained model for ImageNet, which classifies images into 1,000 object categories. Thus, our task is to use \name{} to explain GoogleNet's behavior in the context of instances and classes for the ImageNet dataset.

\subsection{Backend: Concept Extraction and Clustering}
\label{sec:backend}

We use a combination of TCAV~\cite{tcav} and ACE~\cite{ace} as the methodological backbone for concept-based explanation. Combining these methods together lets us automatically derive influential concepts for each class predicted by GoogleNet and the extent of the class' influences. By subsequently clustering the ensemble of derived concepts, we impose a human-understandable structure on the concepts that approximates the model's mental model in classification.

\subsubsection{TCAV}

TCAV, or \textit{testing with concept activation vectors}, is a concept-based explanation method that measures the importance of human understandable concepts to the neural network's inference --- e.g., how much a ``stripe'' concept affects a classifier when predicting a zebra image. The TCAV approach views concepts as a set of images of similar traits. A set of images without these traits forms a set of counterexamples (i.e,. non-concept examples). 

TCAV defines a concept activation vector, or CAV, as the normal to the hyperplane that linearly separates high-dimensional representations of concept examples from that of non-concept examples.
To measure the influence of a concept to an instance's prediction, TCAV computes pixel gradients of the instance at a target layer and compares its direction with the CAV's (both are in the same latent space) by taking inner product of the two. If the gradient vector lines up with the CAV (indicated by a positive inner product) it means the concept is positively influencing the prediction (``stripe'' makes an instance more likely to be a zebra). Conversely, if the two vectors part (indicated by negative inner product), the concept negatively influences the prediction.

The TCAV score, which measures conceptual influence, is defined as: $ TCAV_{Q_{C,K,L}} = \frac{|\{x \in X_k : S_{C,k,l}(x) > 0\}|}{|X_k|}$
where the fraction of \textit{k}-class inputs whose \textit{l}-layer activation vector was positively influenced by concept \textit{C}. $S_{C,k,l}(x)$, is the aforementioned innerproduct of pixel gradients and CAVs. To prevent meaningless concepts (concepts with influence $\sim$0.5), multiple CAVs can be trained using the same set of concept examples against various sets of counter examples. A meaningful concept should lead to TCAV scores that rejects the hypothesis of a 0.5 TCAV score with statistical significance (we use a threshold of $p>0.01$).

In our system, we train 20 CAVs for each concept on concept examples (this parameter can be edited via \name{}'s frontend) and randomly formed counter examples after it is generated using ACE method. TCAV scores are computed for each CAV and averaged.

\subsubsection{ACE}

TCAV provides a mechanism for measuring the influence of human-understandable concepts on the neural network for a specific class, but in the base TCAV approach, forming a set of examples for a concept is a manual process. To automate the process of concept generation and extraction, Ghorbani et al.~\cite{ace} proposed ACE, or \textit{automatic concept-based explanations}.

ACE extracts a set of concepts for each class the model predicts on, by segmenting the images of the class and grouping segments that form clusters in the high-dimensional space of specific latent layer; the TCAV procedure can be followed to filter out meaningless collections. This combined TCAV-plus-ACE process makes efficient and automated concept generation tractable. In our system, we default to sampling 50 images for each class, and segment them using SLIC~\cite{achanta2010slic} with three resolutions (15, 50, and 80 segments) per image. The segments are embedded in the network's ``mixed4c'' layer. $K$-means clustering is then performed to cluster their latent representations into 10 concepts (high-dimensional clusters), though each of these parameters is controllable from \name{}'s frontend.

\subsubsection{Concept Clustering}

Running ACE will extract a large ensemble of concepts. For example, we extracted 1,211 concepts from 143 randomly selected classes in ImageNet using the default parameters mentioned above. This concept space is unorganized (i.e., there's no ranking or structure imposed on them), which makes it difficult to review, explore, and compare them.

An additional consideration (or complexity) is that concepts discovered in different classes might be semantically similar (e.g., ``snow background'' in husky and ``snowy background'' in wolf); such similarity needs to be demonstrated because it is useful for understanding the model from a concept perspective.

To enable structured and guided navigation within this the concepts space (G1), we apply clustering to the extracted concepts. The intent is to group semantically similar concepts into the same cluster, which will let us leverage cluster identity as a means to  demonstrate conceptual overlap across classes, and to form a meaningful navigation structure that the user can rely on to probe and explore the concept space. 

The similarity between two concepts is quantified by their cluster identity (if two concepts come from the same cluster their similarity is 1, otherwise 0). In the frontend, users can choose between $k$-means and agglomerative clustering, and select associated cluster parameters. Based on feedback from our user study evaluation (see Sect.~\ref{sec:user_study}), this provides good flexibility by allowing users to explore different clustering levels and granularities, though the system is extensible to other clustering methods (i.e., as future work) and heuristics. In particular, clustering itself can benefit from the use of XAI techniques, though exploring such activities is beyond the scope of the current paper.

\subsection{Frontend: \name{} Interface}
\label{sec:frontend}
 
Fig.~\ref{fig:teaser} shows the \name{} interface, which consists of four primary panels and several coordinated views (A--N).

\subsubsection{Class Navigation Panel}

The leftmost \textbf{class navigation panel} provides three visualizations (A--C) to facilitate navigation and selection of classes of interest for concept exploration. This panel primarily supports exploring the model from a class-based perspective, and allows users to analyze the concept space at a global level (G1), comparatively analyze classes (G2 and G3), and inspect specific instances within a class (G4).

(A) The \textbf{class performance view} summarizes model performance. Users can begin exploration by brushing histogram bins to filter out undesired classes (e.g., show only low performance classes). 

(B) Such brushing updates the \textbf{class navigation view}. This view provides a global perspective of all classes using t-SNE~\cite{van2008visualizing}, which maps a high dimensional representation of classes (computed as the average of latent vectors of all images of the same class) into a two dimensional embedding. 
To prevent overplotting, we aggregate classes into ``clique'' circles, sized by the number of total classes and colored based on average accuracy of classes in the clique. Hovering on a circle displays a tooltip (see (O)) showing individual classes belonging to the clique with with a representative image per class. Users can select a class of interest for subsequent analysis.

(C) Selected classes are displayed in the \textbf{class confusion matrix view}. The confusion matrix supports inter-class and intra-class analysis by summarizing the classification performance against ground truth labels across the selected classes. 
Clicking on a cell shows a list of images that belong to the corresponding category of the cell below the confusion matrix. For example, Fig.~\ref{fig:teaser}(C) shows the 11 instances of \textit{tiger cat} which were misclassified as \textit{tiger}.

Hovering over an instance shows the concepts that affect the instance's prediction. 
For example, in Fig.~\ref{fig:usage_scenario_3}(J) (part of usage scenario \#3), a \textit{police van} image was misclassified as \textit{ambulance}. Hovering on the instance shows how influential the concepts of these two classes affect the prediction. In the usage scenario, the \textit{ambulance} concepts have slightly higher, influence scores than \textit{police van} concepts, which explains why the model misclassified the image as \textit{ambulance}. 

\subsubsection{Concept Navigation Panel}
\label{sec:concept_navigation_panel}
The \textbf{concept navigation panel} on the right side of the interface contains three views that facilitate navigation of the concept space in a structured way (G1) while providing context during class-concept and instance analysis (G2--G4).

(D) The entire concept space is displayed in a hexagon plot in the \textbf{concept navigation view}. The design of this view underwent multiple iterations during the iterative prototyping process. For example, Fig.~\ref{fig:prototypes}(top) in the Appendix shows an early visualization design of the concept space, based on dimensionality reduction where each concept was plotted as a point. This design had several limitations, including overplotting and difficulty in distinguishing and selecting individual concept clusters.

The hexagon design avoids these issues. Instead of plotting a point cloud, we use the IsoMatch method~\cite{isomatch} to plot organize concepts within a hexagonal grid layout like a heatmap. Each concept is visualized as a hex tile, assigned to a unique location such that (a) clutter and overlap is avoided, (b) cluster boundaries are easily demarcated using dark borders, (c) concepts are evenly distributed across the panel's available space, and (d) concept clusters are easily selectable.

(E) Clicking on a cluster loads it into the \textbf{concept cluster detail view}. This chart summarizes the concepts belonging to a concept cluster. Concepts are ordered along the x-axis by their influence scores, which are mapped to the y-axis. Circles are colored using a diverging blue-to-red color scale based on positive (above 0.5) or negative (less than 0.5) influence score; the influence score mid-point (0.5) is shown as a dotted line, and the average influence score over all concepts in the cluster is shown as a solid horizontal line.

(F) Hovering or clicking on a concept loads it into the \textbf{individual concept view}, which loads all image patches for that concept. By skimming through these, the user can quickly gain a semantic understanding of what that concept is. Skimming multiple concepts in this way can help show why they belong to the same concept cluster. As an additional feature for this view, we provide users with an annotation feature to input descriptive information about a concept cluster. This information is saved for future reference.

\subsubsection{Class-Concept Panel}
The central \textbf{class-concept panel} synthesizes information from the left (class-focused) and right (concept-focused) panels, supporting interpretation to the underlying model's behavior on the given data at inter-class, intra-class and instance levels (G2--G4). It consists of three views, organized around a workflow going from inter-class comparison, to intra-class comparison, to instance analysis.

(G) The \textbf{concept card view} provides an overview of selected classes and concepts (e.g., selecting a set of classes from the class navigation view). Each class is represented by a card, with the class name at top. Below each class title, a histogram shows the distribution of influence scores for the concepts in that class. If the histogram skews to one side, it means the concepts are likewise skewed either positively or negatively influential. In contrast, a normal distribution indicates the majority of concepts in that class are closer to neutral (i.e., close to the 0.5 influence mid-point). 
 
Below the histograms, a detailed dot plot shows individual concepts as circles. Circles are organized into rows corresponding to their concept clusters (labeled at left, e.g., the first row ``CC9'' stands for ``concept cluster \#9''). Circles located on the same row belong to the same concept cluster (i.e., they would be in the same cluster in the concept navigation view in (D)), even when they are distributed among different classes. The left-to-right position of a circle (along with its corresponding red-to-blue color scale) within its class card is based on its influence score along the negative-to-neutral-to-positive scale. 

Concepts in the same row are also linked via horizontal lines to emphasize their concept cluster grouping (supporting inter-class analysis, G3). This design is inspired by Lex et al.'s work~\cite{lex2014upset}. Clicking or hovering on these links highlights the corresponding concept cluster in the hexagon plot in (D), and loads the specific concept cluster in the detail view in (E).

To support additional concept analysis, we adopted the idea of periphery plots~\cite{morrow2019periphery} to supply condensed insights to the left and right of the class cards. (I) The box plots to the left summarize the influence scores for the concepts for each row (i.e., for each concept cluster). (H) To the right, the bar chart shows the number of concepts belonging to that concept cluster, based on the current selection of classes. A higher frequency number indicates that this concept cluster is likely a common characteristic of these classes in the model's mental map.

(J) Clicking the title bar of a class card loads the \textbf{class-specific concept view}. This view  facilitate intra-class concept inspection (G2). Each row shows one concept, sorted based on decreasing influence score (i.e., the top concept rows are more positively influential to the prediction of the class). Users can review and verify if the model is making inferences about a class based on a set of reasonable concepts. For example, if the user sees a concept relating to an image's background, as opposed to important semantic features (e.g., for a \textit{zebra}, ``grassland'' as opposed to ``stripes''), it may indicate that the neural network is assigning too much weight to the background of the image while ignoring foreground information that might reasonably be important (at least from a human's perspective). 

In each row, we present five representative image patches of the corresponding concept so that the user can form a rough idea of what the concept is about. Additional samples can be loaded by clicking the ``+'' icons beside each label.

(K) To the right of the class-specific concept view, the \textbf{instance influence matrix} visualizes local explanations about the models behavior for individual instances. This heatmap plots each row as a concept (aligned with the concepts to the left in (J)), where each column indicates an image sample in the dataset. Thus, each cell indicates, for a particular instance, how influential that concept was in its prediction, colored using the same red-to-blue color scale from previous views. Above each column, a check mark or \textsf{X} indicates if the prediction was correct.

Instances are sorted based on their classification accuracy and confidence (i.e., the left-most column is the correctly classified instance with the highest confidence and the right-most column the misclassified instance with the highest confidence). In Fig.~\ref{fig:teaser}, the user has scrolled all the way to right, showing incorrect predictions with the highest confidence.

(L) Clicking on a column loads its instance into the \textbf{instance view}. This card provides details about the instance (G4), including its image with concept patches that can be toggled on as overlaid semi-transparent boundaries, its ground truth labeling, and the model's prediction and confidence. At the bottom of the card, a lollipop chart shows the concept influences specific to that particular instance.

\subsubsection{Header Bar Panel}
(M) The \textbf{header bar} provides control widgets allowing users to update parameters and re-run the backend pipeline. The menu includes selecting a dataset and model, selecting the model layers where the concepts are extracted, setting TCAV and ACE parameters, specifying the concept clustering methods and number of clusters (i.e., the parameter $k$ in $k$-means). The number of clusters is set by default to the optimal value based on the silhouette score~\cite{rousseeuw1987silhouettes}. 
Users can hover on other numbers of clusters to see their scores as shown in (M).

\newcommand{\person}{Michael}

\section{Usage Scenarios}
To help demonstrate how \name{} supports concept-based explanation to learn about the model behavior, we present three usage scenarios exploring GoogleNet on the ImageNet dataset. Each focuses on a different type of exploration and analysis: (\#1) for individual classes and their instances,  (\#2) comparing the concepts between two classes, and (\#3) understanding concept commonalities across many classes. We tell these stories from the perspective of \person{}, a non-expert in AI.

\begin{figure}[h]
    \centering
    \includegraphics[width=\columnwidth]{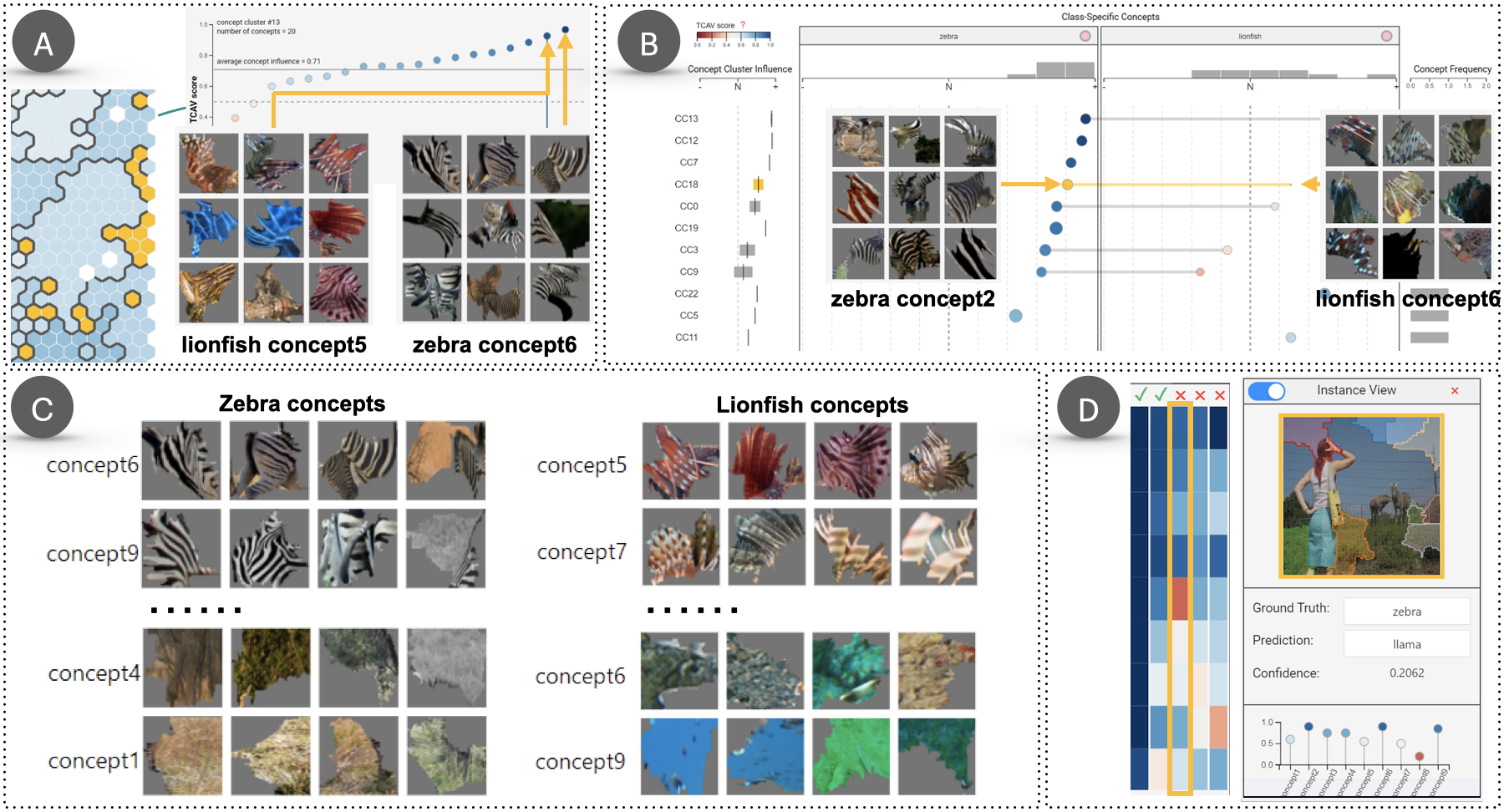}
    \caption{Usage scenario \#1 represents analysis of the \textit{zebra} and \textit{lionfish} classes. A full-size figure is available in the Appendix.}
    \label{fig:usage_scenario_1}
    \vspace{-.5cm}
\end{figure}

\subsection{Usage Scenario \#1: Verifying GoogleNet's knowledge about individual classes and samples}

As Fig.~\ref{fig:usage_scenario_1}(A) shows, \person{} begins by reviewing clusters in the concept navigation view, focusing on clusters with darker colors (i.e, containing more influential concepts). Loading Cluster \#13 in the concept cluster detail view, he realizes the two most-influential concepts (from the \textit{zebra} and \textit{lionfish} classes) look like stripes. He loads the \textit{zebra} and \textit{lionfish} classes into the central class-specific concepts panel for subsequent investigation.

In the class-specific concepts panel, \person{} sees several cross-class links between the \textit{lionfish} and \textit{zebra} classes, which further confirms that conceptual overlap exists (Fig.~\ref{fig:usage_scenario_1}(B)). Reviewing these, he sees that the shared concepts primarily relate to stripe patterns. As both zebras and lionfish have prominent stripe patterns, GoogleNet has learned (or formed a mental model) that stripe patterns are a way to recognize both zebras and lionfish.

\person{} next inspects the individual concepts of the \textit{zebra} and \textit{lionfish} classes (Fig.~\ref{fig:usage_scenario_1}(C)).
For the \textit{zebra} class, the three most positively influential concepts pertain to stripe patterns. As influence decreases, the concepts become more associated with noise and background features, such as representing desert or grassland. In this case, GoogleNet is aligning with \person{}'s mental model, as background information is largely irrelevant to classifying these images as zebras.

Reviewing individual samples using the influence instance matrix and the instance view, \person{} notices a misclassified image of \textit{zebra} predicted as the \textit{llama} (Fig.~\ref{fig:usage_scenario_1}(D)). The zebras in this image are very small, and ``background concepts'' (such as grasses and sky) are the main concepts present. In other words, the model predicted using less-influential concepts that were present, as stripe concepts were not found. This happens several times in instances predicted as \textit{zebra} or \textit{lionfish}, and the model regularly predicts wrongly in such cases.

\begin{figure}[tbh]
    \centering
    \includegraphics[width=\columnwidth]{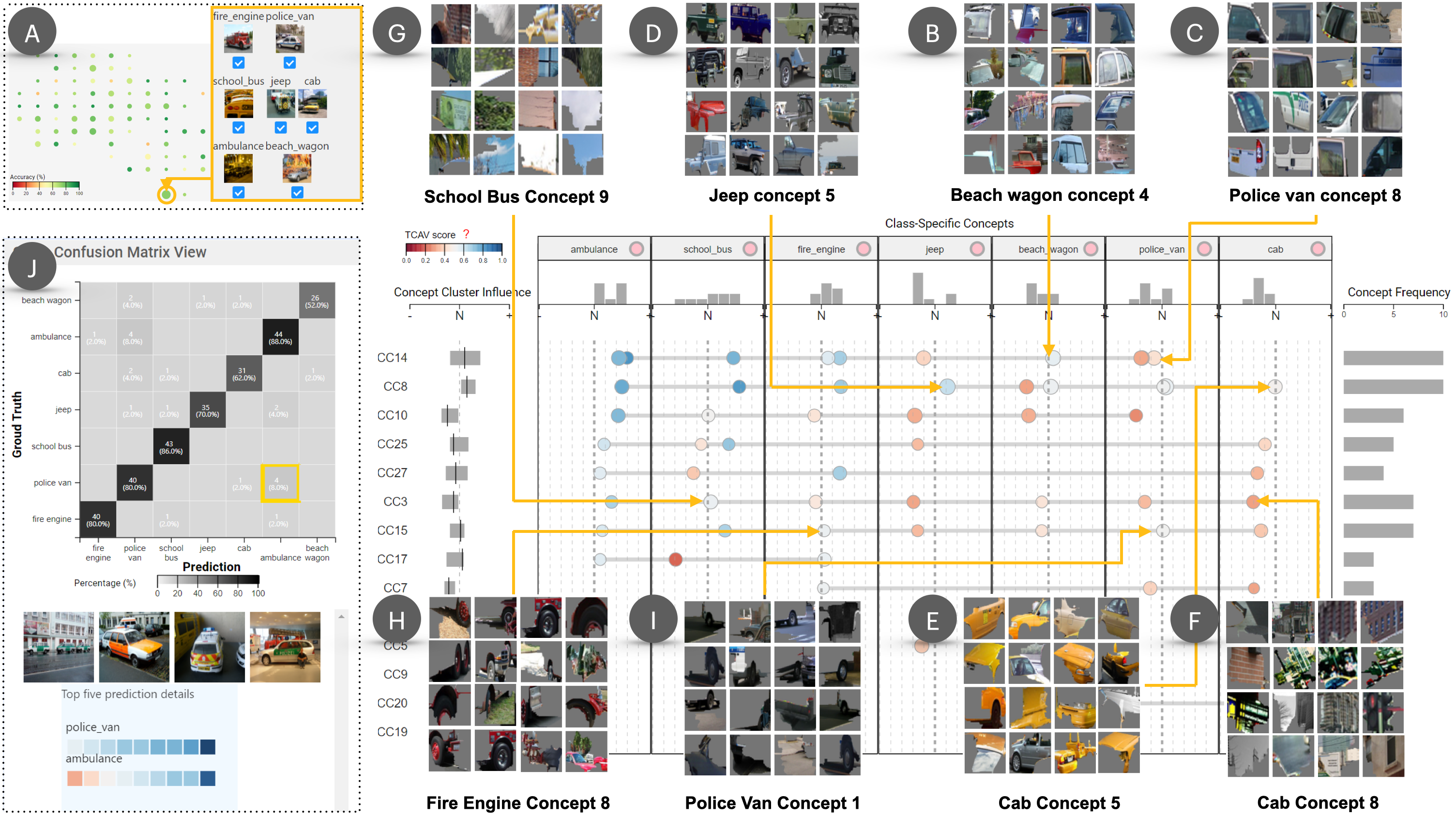}
    \caption{Usage scenario \#3 represents analysis on a group of classes. A full-size figure is available in the Appendix.}
    \label{fig:usage_scenario_3}
    \vspace{-.5cm}
\end{figure}

\subsection{Usage Scenario \#2: Spotting unreasonable concepts and data quality issues}
\label{sec:usage_scenario_2}


Using the class navigation view, \person{} looks at other classes similar to \textit{zebra}. Hovering on the circle containing the \textit{zebra} class displays a tooltip, which indicates the \textit{tiger} class (another animal with prominent stripes) is also contained in this clique (Fig~\ref{fig:teaser}(O)). An adjacent clique in this view is colored in red (indicating below-average classification accuracy); the tooltip shows that this clique contains a \textit{tiger cat} and \textit{tabby} class. Intrigued, he selects the \textit{tiger}, \textit{tiger cat}, and \textit{tabby} classes. 

The class confusion matrix (Fig.~\ref{fig:teaser}(C)) view shows many misclassifications between \textit{tiger cat} and \textit{tiger}; for example, there are 11 images of \textit{tiger cat} (ground truth) that were predicted as \textit{tiger} by GoogleNet. Selecting this cell loads these images; to \person{}'s eyes, they all indeed appear to be \textit{tigers}, not \textit{tiger cats}. Already, \person{} can see that there is a data labeling issue in ImageNet: instances that should have been labeled as \textit{tigers} were wrongly labeled as \textit{tiger cats}.

Most \textit{tiger cat} concepts appear to have negative influences in concept card view (Fig.~\ref{fig:teaser}(G)). \person{} reviews these in the class-specific concept view (Fig.~\ref{fig:teaser}(J)). 

He quickly scans through the concepts, and unsurprisingly, many of them seemed to be random or incomprehensible patches that do not form a collective theme. 
In other words, GoogleNet did not form a sensible mental model about \textit{tiger cat}.
Interestingly, the most negatively influential concept at the bottom row presents a ``tiger stripe'' pattern,
which helps explain why there are so many \textit{tiger cats} misclassified as \textit{tigers}. 

Because of this, \person{} decides to review \textit{tiger cat} instances that contain the ``stripe'' concept using the instance influence matrix and the instance view (Fig.~\ref{fig:teaser}(N)). 
Reviewing misclassified instances, he again can see that there are several images labeled as \textit{tiger cat} but actually including \textit{tigers} in them. 
Overall, \person{} observes that mislabeled instances tend to include small objects of tigers. 
This explains why the tiger stripe was not positively influential -- because it was not easily detected due to its size.

\subsection{Usage Scenario \#3: Understanding the neural network's knowledge on a group of classes}


In Fig.~\ref{fig:usage_scenario_3}(A), \person{} investigates GoogleNet's global mental model by reviewing the class navigation view. He notices a big circle in the bottom right corner of the visualization, indicating a clique containing many classes. The tooltip shows that the classes (7 in total) are all vehicles. He selects all of them.
Reviewing these in the class-specific concepts view, he sees (as expected) many cross-class links in the class-specific concepts panel which connect influential concepts for the various vehicles. 
This means the neural network has likely learned common features across different cars.

\person{} scans through the horizontal links across classes one by one. Fig.~\ref{fig:usage_scenario_3}(B) and (C) show that the highest-ranked concept cluster (labeled CC14, for concept cluster \#14) contains window-like patches in the \textit{beach wagon} class, so an important ``vehicle commonality'' the network has likely learned is ``car windows.'' 
This hypothesis is confirmed by reviewing across other concepts in this concept cluster (e.g., Fig.~\ref{fig:usage_scenario_3}~(C) also contains window patches from the \textit{police van} class). The next set of links in Fig.~\ref{fig:usage_scenario_3}~(D) and (E) tend to show ``car side'' concepts, which are relevant features regardless of the vehicle type (in this case, between \textit{jeep} and \textit{cab}. Further investigation reveals both ``urban background'' in Fig.~\ref{fig:usage_scenario_3}~(F) and (G), and ``wheel'' as influential concepts in Fig.~\ref{fig:usage_scenario_3}~(H) and (I). Both concepts make sense at a high level (``\textit{Cars have wheels and car pictures are usually taken in urban environments.}''), but interestingly, the \textit{jeep} class does not consider the ``urban background'' concept as influential. 
Reviewing \textit{jeep} instances reveals why: many \textit{jeep} pictures are taken with the background in nature (desert or woodland settings); he cannot find \textit{jeep} pictures taken in urban settings.

While many of the concepts make sense upon review, \person{} is surprised to realize that GoogleNet considers ``windows'' and ``car sides'' as the \textit{most salient} discriminators or influencers for predicting cars, while other features like ``wheels'' or ``headlights'' are not chosen.

\section{User Study}
\label{sec:user_study}

To evaluate \name{}, we designed a user study to answer two primary questions relating to design goals (``\textit{How well does the system support the design goals (G1--G4)?}'') and overall usability (``\textit{What is the overall usability of the system?}''). Our user study consisted of two stages: i) a task stage where participants completed three representative analysis tasks in line with the four design goals listed in Sect.~\ref{sec:design_goals}; and ii) a freeform analysis stage where participants freely probed model behavior using the interface. We recruited ten participants who were non-experts in deep learning, had not heard of concept-based explanations, and had not used ImageNet and GoogleNet before. We collected both quantitative and qualitative data, which allowed us to robustly evaluate both the ability of \name{} to support the design goals G1--G4, and also to understand the system's overall usability.

\subsection{Study Design} 

Participants took the following procedure for the study:

\textbf{(1) Training.}
Participants first completed a short demographics questionnaire. Next, they were given a brief (high-level) introduction on TCAV, ACE, and what concept influence means. The study administrator then walked the participant through available features and functionalities of \name{}; participants completed a simple training task to help familiarize themselves with the system. During the training, participants could ask questions at anytime, and were allowed to play around with the interface until they felt comfortable to proceed.

\textbf{(2) Task.} Participants completed the following three tasks (T1-T3):

\textbf{T1:} Inspect the influential concepts for a given class: (1) Identify the comprehensible concepts and state why they are comprehensible. (2) Identify the incomprehensible concepts and state why they are incomprehensible. (3) Review the concept influences; what are the items that make sense to you, and what are the items that do not make sense to you? (4) Take a look at the instances contained in the influence matrix, find interesting cases and describe why they are interesting. This task primarily focuses \textit{instance-level} and \textit{single class-level analysis}, and supports G1 and G2. Each participant completed T1 on two classes: one high-performance (accuracy $>0.9$) and one poor-performance (accuracy $\sim0.1-0.3$); class choices for each participant were randomly selected from a high-performance set (\textit{zebra}, \textit{tench}, \textit{lionfish}) and a poor-performance set (\textit{tiger cat}, \textit{appenzeller}, \textit{seashore}).

\textbf{T2:} Given two classes that are similar and have overlapping misclassifications (i.e., images which should belong to $C_2$ are classified as $C_3$, and vice versa): (1) Identify what are the commonalities between their influential concepts. (2) Identify distinguishable influential concepts unique to each class. This task represents \textit{instance-level} and \textit{multi-class analysis}, and supports G2 and G3. The class pairs for T2 were randomly selected from the following sets: \{\textit{tiger cat}, \textit{tiger}\}, \{\textit{eskimo dog}, \textit{siberian husky}\}, and \{\textit{police van}, \textit{ambulance}\}. Note that if \textit{tiger cat} was chosen during T1, it was not an option in T2.

\textbf{T3:} Given a group of seven related classes representing various vehicles (\textit{fire engine}, \textit{police van}, \textit{school bus}, \textit{jeep}, \textit{cab}, \textit{ambulance}, \textit{beach wagon})): 
(1) What are the common influential concepts across this group? Reason about the vehicle knowledge that the neural network has learned and what vehicle knowledge it lacks. This task represents \textit{multi-class} and \textit{global-level analysis}, and supports all four design goals (G1--G4). All participants used the same set of seven classes for this task. (Two participants had previously seen the \textit{police van} and \textit{ambulance} classes in T2, however their results were in line with other participants, so we do not believe this caused any study confounds during this task.)

The task order was consistent. Each participant was given a verbal description of the task (which was also available on a sheet of paper) by the administrator. The think-aloud protocol was employed during this stage; the administrator listened to verbal utterances to help confirm that the participant was correctly performing the task and to check if the participant's interpretations of tasks and system features was correct. Upon completion of a task, the participant verbally summarized their answer(s) to the administrator.

\textbf{(3) Freeform analysis.} In this stage, participants conducted an undirected, freeform analysis of ImageNet to gain insights on model behavior using \name{}. Participants were told to use the tool until they were satisfied, but were required to spend at least 10 minutes. To prevent participants from becoming lost or frustrated, we prepared an initial (optional) motivation scenario which was only given if participants asked for guidance: ``Check if the neural network is working sensibly on well performing classes.'' Only one participant asked for this, as we found that the others had ideated their own exploratory goals upon completing the task stage. Participants also verbally reported their thought processes during this stage.

\textbf{(4) Review.} Participants completed a short usability survey to rate various system components using 7-point Likert scales; they were also allowed to provide comments or critiques about the system.

\subsection{Participants and Apparatus} We recruited ten participants: nine graduate computer science students at Arizona State University and one analyst with two years experience in a data company (average age = 25.8, SD = 2.08; 8 males, 2 females). Although a couple of the participants had colloquial familiarity with AI/ML, none had expertise in deep learning development or analysis, and all were unfamiliar with concept-based explanation methods. Study duration averaged 86 minutes (SD = 15).

\name{} was shown using Google Chrome in full-screen mode on a 30'' monitor at $3840\times2160$ resolution. Study sessions were conducted in a quiet, office environment with no distractions.

\subsection{Study Results}

To analyze the study, we first report quantitative ratings from the review stages' usability survey. We next report on qualitative verbal comments and responses given by participants, which were collected via the think-aloud protocol and summary answers during their tasks.

To analyze these verbal comments, we used a grounded theory procedure~\cite{thornberg2014grounded} to qualitatively code comments based on assessing how \name{} promotes insights and supports the design goals G1--G4.

\begin{figure}[t]
    \centering
    \includegraphics[width=.9\columnwidth]{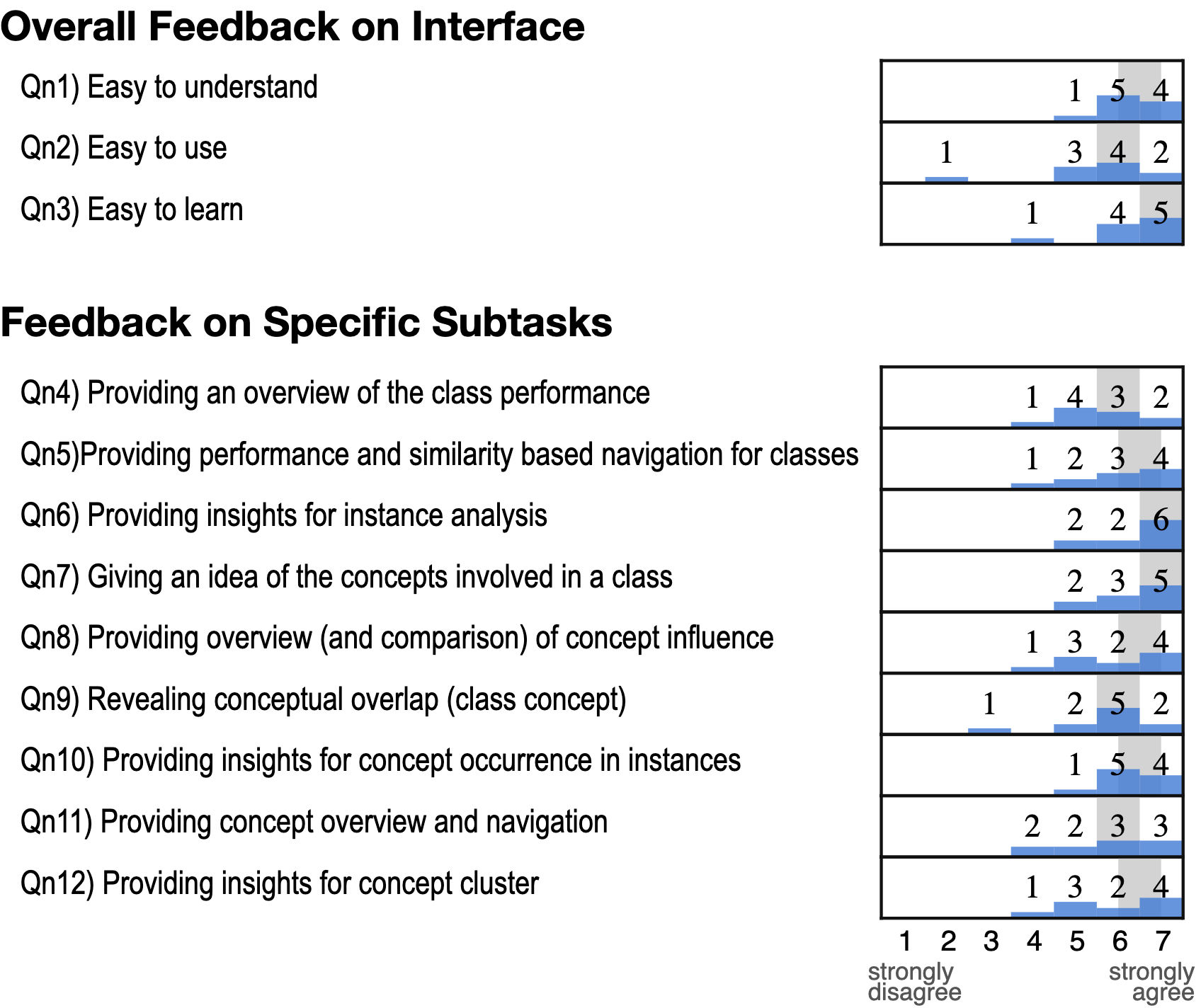}
    \caption{Participant  usability ratings about \name{} based on the survey given during the user study's review stage. Median ratings are indicated in gray. }
    \label{fig:study_ratings}
    \vspace{-.5cm}
\end{figure}

\subsubsection{System Usability Ratings}

Fig.~\ref{fig:study_ratings} summarizes system usability based on participants ratings from the survey completed during the review stage. Ratings are broken into two types: (Q1--Q3) the overall usability and effectiveness of the system, (Q4--Q13) the usability of individual features. 

In general, feedback was positive. We highlight that the system was considered easy to use (Q1), learn (Q2), and understand (Q3), and that the individual interface panels, the class navigation panel (Q5--Q6), the class concept view (Q7--Q11), and the concept navigation panel (Q12--Q13) were all positively regarded.

\subsubsection{Feedback from Novice Users}

The survey ratings show that \name{} achieves good usability. Here, by analyzing the verbal comments and responses of participants, we reflect on the types of insights users can generate with the system, specifically in the context of the design goals (G1--G4).

\textbf{(G1) Contextualizing concepts using the concept navigation view.} The concept navigation view was heavily used and considered positively as a way to explore new concepts. In particular, several participants inspected similar concepts using thi view as a way to verify their intuition or mental model about a concept of interest. ``\textit{I'm trying to see what are concepts similar to this one because I want to know if my thought is correct}'' (p5). ``\textit{This map thing [the concept navigation view] is cool. I can compare this [concept] with similar ones}'' (p8). ``\textit{The navigation view helps me understand a concept better}'' (p9).

\textbf{(G2) The concept view is effective revealing class-specific concepts.} Several participants regarded the class-specific concept view within class-concept view as effective in revealing influential concepts. Specifically, the list of concepts ranked by concept influences provided them with intuitive insights for network behaviors. ``\textit{This is helpful because I can immediately see what the neural network learned about the class}'' (p3). ``\textit{It doesn't make sense that top concepts are background patches...middle concepts should be moved up. It's clear in the view}'' (p8).

Despite liking the class-specific concept view, participants sometimes encountered incomprehensible concepts during their analysis. While most concepts had ``good quality,'' and participants could understand them at a glance, some other concepts did not present a uniform pattern, which sometimes led participants to get stuck. Trying to make sense of such concepts seemed infeasible. In part, this is a byproduct of the concept generation pipeline used in \name{}'s backend, which automatically extracts and defines concepts (bottom up) as opposed to creating them by hand using some rules (top down). 
Unfortunately, this process is not error-prone as it can extract and define noisy, unhelpful concepts, too. When users encounter with such concepts, they can slow down their analysis process. We discuss this issue more in Sect.~\ref{sec:user_study_takeaways}.

\textbf{(G3) Concept links are effective revealing inter-class concept overlap.} Participants reported it was easy to identify conceptual links between classes and understand them by using the concept view. ``\textit{I can see that concept cluster 5 are green areas in the two classes.}'' (p2, analyzing the \textit{tiger cat} and \textit{tiger} classes), ``\textit{The first link is white-fur stuff.}'' (p5, analyzing the \textit{Eskimo dog} and \textit{Siberian husky} classes), ``\textit{I can see silhouette of cars, tire and window [being the common concepts].}'' (p7, analyzing the \textit{police van} and \textit{ambulance} classes). As an extension of the current system, two participants (p1, p9) requested the ability to compare two or multiple concepts in parallel.

\textbf{(G4) Instance analysis was found to be useful for identifying data quality issues.} 

A couple of participants identified (unprompted) data quality issues solely using instance analysis view. ``\textit{Clearly, real \textit{tigers} are mislabeled as \textit{tiger cats} in the testing set}'' (p3). ``\textit{Why is there a tiger stripe in \textit{tiger}\_cat\_concept\_8? It doesn't make sense ... Oh I see why, because there are tiger pictures in ground truths}'' (p8).

Although participants in general like the instance analysis views, four participants (p1, p5, p6, p8) mentioned it was hard for them to link concepts with their segments in the image instance view. Three participants (p1, p3, p7) mentioned that instances sometimes seemed counterintuitive. ``\textit{Some instances are highly (positively) influenced by the concepts but are still misclassified, why?}'' (p7, while analyzing \textit{tiger cat}). ``\textit{Why are `police' letters negatively influential  here?}'' (p3, analyzing \textit{police van}). ``\textit{Why are cab cases all negatively influenced but correctly classified?}'' (p1, comparatively analyzing multiple car classes). The current interface does not support answering these sorts of ``\textit{why?}'' questions; see Sect.~\ref{sec:user_study_takeaways} for discussion on this.

\section{Discussion and Conclusion}
\label{sec:discussion_conclusion}

We have presented \name{}, an interactive visual analytics system that supports non-expert users to explore and probe concept-based explanations of deep learning models. Based on a set of usage scenarios and a robust user study, we demonstrate that \name{} addresses a number of design challenges (C1--C4) and goals (G1--G4) that are important to the problem of concept-based explanation for non-expert users. In particular, we show how non-experts can effectively reason about model behavior at different analytic scales (i.e., instance, class, and global levels). Below, we discuss takeaways and lessons learned from our experiences in designing and evaluating \name{}, which include identifying current system limitations and opportunities for future research directions.

\subsection{User Study Takeaways}
\label{sec:user_study_takeaways}
In general, the user study demonstrated the overall performance of our system from a human-centric perspective, indicating that it provides good usability and successfully supports the design goals (G1--G4), while also illustrating nuances and complexities in concept-based explanations. For example, some concepts in the system were not comprehensible to participants, which could slow down (or derail) their analytic process. Unfortunately, there is no easy way around this issue; in fact, it is an acknowledged problem that automatic concept discovery invariably introduces some amount of noise~\cite{ace}. As a future workaround, \name{} employs a modularized design. As novel concept discovery methods are developed (which hopefully reduce noise), they can be integrated into our system.

Another interesting takeaway from our study is that participants
tended to use the instance analysis view more than we expected; based on feedback, we see several potential avenues for future extensions to enrich functionality. For example, the current instance analysis view provides instance-level explanations in terms of individual concept influences. A logical next step could be illustrating how the interplay of different concepts influence an instance's prediction. For example, both ``snowy background'' and ``white fur'' are positively influential for the \textit{white wolf} class. When these two concepts co-occur in an instance, do they increase the likelihood of a \textit{white wolf} prediction, compared to if only one concept is present? To achieve this type of fine-grained analysis, we plan to adopt techniques such as VRX~\cite{vrx} in future versions of \name{}.

\subsection{Serendipitously Supporting an Unexpected Task}

Our user study also revealed an explanatory task supported by \name{} that we did not intentionally support. During the freeform analysis stage, several participants analyzed if the model ``understood'' a class by quickly tabbing through instances in the class-specific concept view ((J) in Fig.~\ref{fig:teaser}). This allowed them to comprehend a rough estimate of the types of instances making up the class, and to use this as a basis to understand the influential concepts that the model has learned for the class. This ``fact-checking'' action was unexpected to us, but demonstrates the flexibility of the system's visual analytics to support diverse actions to interpret deep learning behavior.

\subsection{\name{}'s Suitability for Expert Users}

To understand \name{}'s suitability for \textit{expert} users, we conducted a set of pair analytic sessions~\cite{arias2011pair} and semi-structured interviews with three deep learning experts (Ph.D. students with $3+$ years experience in AI research). The intent was to formatively assess the system's potential applicability for expert users, and to investigate how the system could be extended to better support them.

Each opined that the system was easy to understand once the concept method was illustrated and a walk through of the interface was given. 

In terms of altering the system to support expert usage scenarios, it was suggested that concept explanations could be juxtaposed with visualizing the inner architecture and logic of the neural network. Such ``opening up the black box,'' when juxtaposed with concepts, could enable novel ways to investigate and understand peculiar or unexpected model behavior. For example, when training a model, tools like \name{} can be used to identify the conceptual root cause of misclassifications between two similar classes, or to identify issues in ground truth labeling. Along these lines, one expert also suggested we augment the interface to let users focus on revealing concept distinguishability, so it becomes more explicitly clear on what basis the neural network discriminates between classes. In the future, we plan to explore ways to integrate techniques that combine concepts with visualizing the model's inner architecture, though such designs are likely unsuitable for non-experts.

\subsection{Concept-based Explanations in Other Domains}

Currently, concept-based explanations have primarily been applied in image classification scenarios. Likewise, our system is optimized for this type of deep learning task, as opposed to other domains (e.g., natural language processing). As a future step, we would like to explore the use of visual analytics as a modality for probing and exploring concept-based explanations applied to other (currently unexplored) domains, such as speech recognition or time series prediction. Such domains bring their own complexities; e.g., in image classification, a user can directly observe the pixels in a set of image patches representing a concept, but there are no direct analogs to this process in many other domains. Despite this, visualization may prove to be a key approach for this, due to the power of visualization in being able to graphically render abstract data in ways that reveal patterns and insights.

\bibliographystyle{abbrv-doi}

\bibliography{template}
\end{document}